# $^{19}$B isotope as a $^{17}$B-$n$-$n$ three-body cluster close to unitary limit

**J. Carbonell**[1], **E. Hiyama**[2,3], **R. Lazauskas**[4], **F. M. Marqués**[5]

[1] Université Paris-Saclay, CNRS/IN2P3, IJCLab, 91405 Orsay, France
[2] Department of Physics, Kyushu University, Fukuoka, 819-0395, Japan
[3] RIKEN Nishina Center, Wako 351-0198 Japan
[4] IPHC, IN2P3-CNRS/Université de Strasbourg BP 28, F-67037 Strasbourg Cedex 2, France
[5] LPC Caen, Univ. Normandie, ENSICAEN, Univ. Caen, CNRS/IN2P3, 14050 Caen, France

E-mail: `carbonell@ipno.in2p3.fr`,

**Abstract.** We describe $^{19}$B in terms of a $^{17}$B-$n$-$n$ three-body system, where the two-body subsystems $^{17}$B-$n$ and $n$-$n$ are unbound (virtual) states close to the unitary limit. The energy of $^{19}$B ground state is well reproduced and two low-lying resonances are predicted. Their eventual link with the Efimov physics is discussed. This model can be extended to describe the recently discovered resonant states in $^{20,21}$B.

## 1. Introduction

The S-wave neutron-Nucleon interaction ($n$N) is in overall attractive in all the spin and isospin channels. For the $np$ case, the triplet state ($^3$S$_1$) is more attractive than the singlet one ($^1$S$_0$) and – together with the tensor coupling – it is responsible for the deuteron bound state, while the singlet one remains virtual although very close to threshold. For the $nn$ case, the $^1$S$_0$ potential is very close to the $np$ one and has also a nearthreshold virtual state. These potentials are displayed in figure 1 (left panel) in a simple model, together with the poles of the corresponding scattering amplitudes in the complex momentum plane (right panel). The spin-dependence of $V_{nN}$ manifests in a 20% variation in the attractive strength of the $^3$S$_1$ versus $^1$S$_0$ potentials.

Despite all $V_{nN}$ are attractive, a low energy $n$ scattering on a nucleus (A) feels the other $n$'s in the target and will soon behave – in fact already starting by deuteron – as if the $n$A potential $V_{nA}$ was repulsive. The Pauli principle – imposing an antisymmetric wave function – acts *as if* there was a repulsive interaction among $n$'s. This has dramatic consequences in the 3$n$ and 4$n$ systems: the 3$n$ Hamiltonian, has a ground state bound by $\approx$ 1 MeV but this state is symmetric in particle exchange and not realised in Nature. The same happens with the 4$n$ hamiltonian [1, 2, 3, 4] with a symmetric ground state bound by $\approx$ 5 MeV. As we will see in what follows, the Pauli principle plays also a relevant role in understanding the main features of the low energy scattering of $n$'s on light nuclei.

## 2. n-A scattering length

An interesting observable measuring the repulsive or attractive character of the interaction is the scattering length $a_S$, which can be defined as the $n$A scattering amplitude at zero energy $a_S = -f_{nA}(E = 0)$. Following this convention, a purely repulsive potential generates always positive scattering lengths. For a purely attractive one, its sign depends on the strength of the potential: it starts being negative in a weak potential but changes its sign after the appearance of the first bound state. This behaviour is illustrated in figure 2 in the case of a square well potential.

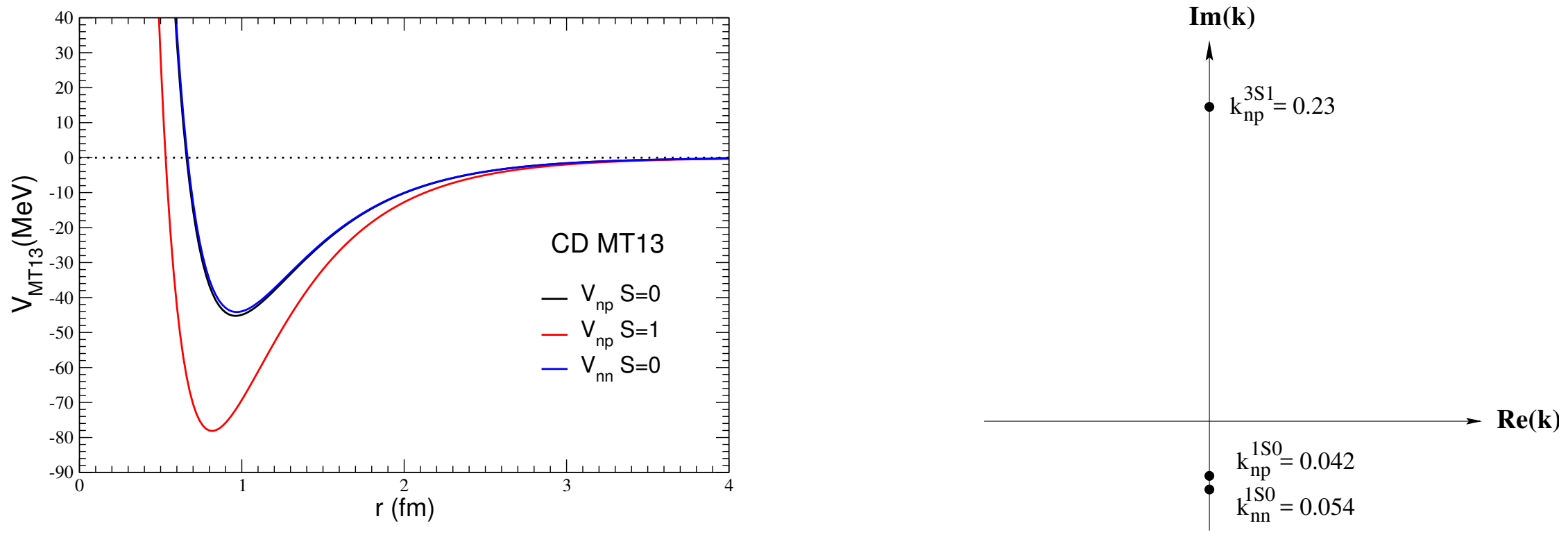

**Figure 1.** S-wave $V_{nN}$ in the different spin an isospin channels (left) and singularities of the corresponding scattering amplitude in the complex momentum plane (right).

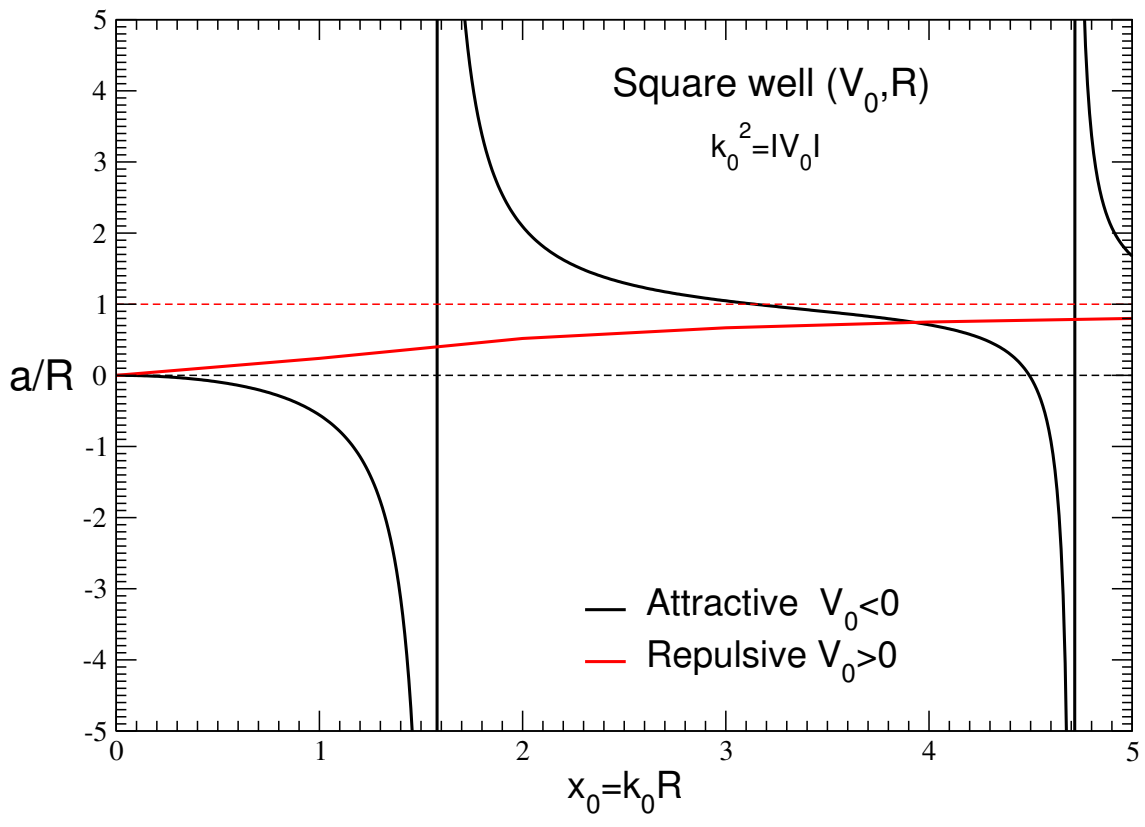

**Figure 2.** Scattering length in a square well potential with parameters $(V_0, R)$ as a function of the strength parameter $x = \sqrt{V_0}R$. Solid red curve corresponds to the repulsive case and solid black one to the attractive case. The later displays the pole singularities due to the appearence of a new bound state in the system.

The evolution of the $nA$ scattering length $a_{nA}$ when the neutron number (N) in the target is increased is displayed in Table 1. The notation $a_\pm$ corresponds to the total spin coupling $S{=}J_A \pm 1/2$, keeping $a_-$ for the case $J{=}0$. Most of the experimental values in Table 1 are taken from the compilations [5, 6]. In some cases, incompatible results are assigned to the same reaction, like for instance for the n-$^3$H, and we have chosen the "Recommended value" or our personal conclusion guided by some theoretical input (see [7, 8, 9] for a discussion). In other cases, the quoted value comes from a theoretical analysis of experimental data [12, 13].

With **N=0** (i.e. A=$p$) both scattering lengths $a_\pm$ are attractive, as expected from figure 1. The positive value of $a_+$ in the np S=1 channel (denoted with an asterisk +5.42*) is due to the existence of the deuteron bound state and so it is attractive, in agreement with figure 2.

Apart from the Pauli forbidden $nn$ $^3$S$_1$ state, the first repulsive channel already appears at **N=1**, in the S=3/2 $n$-$^2$H and $n$-$^2$He (quartet) states ($a_+$). In the shell model representation of the compound system, the two neutron spins must be aligned and occupy s-wave orbitals. what is forbidden by the Pauli principle. In nature, this schematic representation manifest as a repulsive state. Notice that he S=1/2 (doublet) states of these systems – corresponding to antilaligned neutron spins – are not affected by the Pauli principle and are naturally attractive in both systems despite its positive value corresponding to triton $^3$H ($a_-$=0.65* fm) and $^4$He ($a_-$=6.6*-3.7i) bound state. In the later case a negative imaginary part due to its coupling to the $p$-$^3$H channel.

| N Z A | Symbol | $J_A^\pi$ | $a_-$ | $a_+$ | Ref. |
|---|---|---|---|---|---|
| 0 1 1 | p | $1/2^+$ | -23.71 | +5.42* | [5, 6] |
| 1 0 1 | n | $1/2^+$ | -18.59 | ∅ | [5, 6] |
| 1 1 2 | $^2$H | $1^-$ | +0.65* | +6.35 | [5, 6] |
| 1 2 3 | $^3$He | $1/2^+$ | +6.6* -3.7i | +3.5 | [5, 6] |
| 2 1 3 | $^3$H | $1/2^+$ | +3.9 | +3.6 | [8] |
| 2 2 4 | $^4$He | $0^+$ | +2.61 | | [5, 6] |
| 3 3 6 | $^6$Li | $1^+$ | +4.0 | +0.57 | [5, 6] |
| 4 3 7 | $^7$Li | $3/2^-$ | +0.87 | -3.63 | [5, 6] |
| 6 2 8 | $^8$He | $0^+$ | -3.17 | | [10, 11] |
| 6 3 9 | $^9$Li | $3/2^-$ | ≈-14 | | [10] |

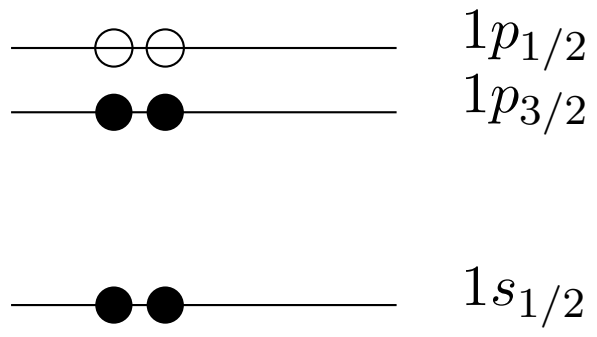

**Table 1.** In left panel experimental nA scattering length (fm) for light nuclei: we denote by $a_\pm$ the S=J$_A$±1/2 total spin state, keeping $a_-$ for the J$_A$=0 case or unassigned values of S (like $^9$Li). Right panel represent the filled shell model neutron orbitals in $^7$Li.

With **N=2** ($n$-$^3$H and $n$-$^4$He) all the scattering length are repulsive. The same happens for the unique **N=3** state: $n$-$^6$Li. It is worth noticing that this strong repulsion – manifested from $^2$H to $^6$Li - manifest only in S-wave. Indeed, most of these systems have $n$A attractive P-waves, which in some cases like $^3$H [14] and $^4$He [15, 16, 17] are even resonant.

For **N=4**, the $n$-$^7$Li scattering length becomes attractive again. Two of the four $n$'s in the target are in P-wave orbital, as illustrated in the right panel of Table 1. As a consequence, the effects due to antisymmetrization with the S-wave incoming $n$ is weakened and the balance between the attractive $nN$ interaction and the Pauli repulsion results in favour of an attractive S-wave state. This attraction persists in $^{12}$Be [12] and $^{15}$B [13].

When arriving at the $^{17}$B, the balance between attraction and repulsion is so fine-tuned that the scattering length becomes huge, indicating the presence of a $^{18}$B virtual state extremely close to the $n$-$^{17}$B threshold. This virtual state was first found in the MSU experiment [18], where the best fit to the data provided an $n$-$^{17}$B scattering length of $a_S = -100$ fm, which constitutes the absolute record of the whole nuclear chart [6]. However a $\chi^2$ analysis of their results allow them to fix only an upper limit $a_S < -50$ fm. Since the ground state of $^{17}$B is a J$^\pi$=3/2$^-$ state the total S of the $n$-$^{17}$B state can be $S$=1$^-$ or $S$=2$^-$ and a shell model calculation indicated that the measured scattering length corresponds to $S$=2$^-$. The MSU results have been recently confirmed at RIKEN [19].

This remarkable experimental finding of an extremely resonant $n$-$^{17}$B system, motivated us to model the S-wave $n$-$^{17}$B interaction and attempt to describe $^{19}$B – also experimentally known – in terms of a 3-body double resonant $^{17}$B-$n$-$n$ cluster. The first results, published in [20], will be sumarized in the following section.

## 3. Modeling $^{19}$B as a $^{17}$B-$n$-$n$ three-body cluster

We have first modelled the resonant S-wave $n$-$^{17}$B interaction by a sum of an attractive term to account for all the V$_{nN}$ attraction and a repulsive one to account for the Pauli repulsion among the incoming and target n's. We have assumed the following form:

$$V_{n^{17}B}(r) \;=\; V_r \left(e^{-\mu r} - e^{-\mu R}\right) \frac{e^{-\mu r}}{r} \tag{1}$$

where $R$ is a hard-core radius, fixing the penetration of the incoming neutron in the target nucleus, and $\mu$ is the range parameter of the effective $n$-$^{17}$B potential. We have fixed $R = 3$ fm, in agreement with the measured r.m.s. matter radius of $^{17}$B [21], and $\mu = 0.7$ fm$^{-1}$ corresponding to the pion mass. At this level we work under the hypothesis of spin-independent $n$-$^{17}$B interaction, in particular both scattering length are equal $a_{1^-} = a_{2^-} \equiv a_S$. In this case, the only remaining free parameter is the strength $V_r$ which is adjusted to reproduce $a_S$ and so the virtual state. Since the precise value of $a_S$ is not know we have considered a wide range of variation. The dependence of $a_S$ on $V_r$, together with the corresponding potential for $a_S = -50, -100, -150$ fm, are displayed in figure 3.

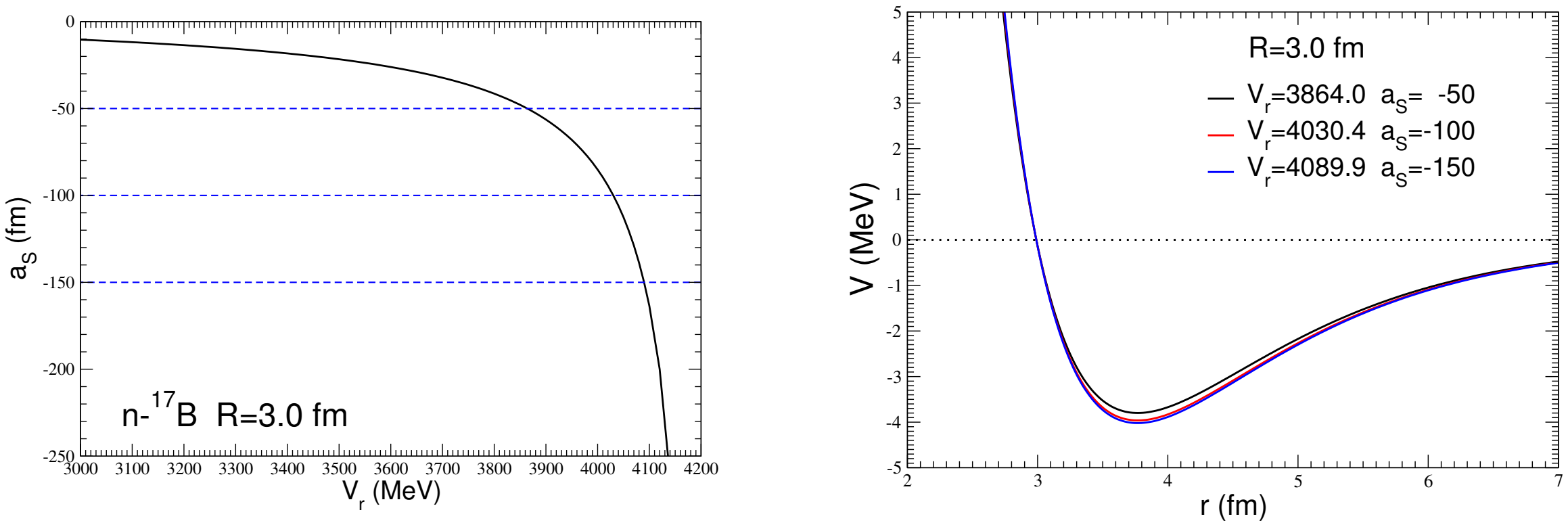

**Figure 3.** Dependence of the $n$-$^{17}$B scattering length $a_S$ on $V_r$ for $R = 3$ fm (left panel). The dashed lines correspond to some selected values of $a_S$=-50,-100,-150 fm, for which $V_{n^{17}B}$ have been drawn (right panel) .

In order to describe $^{19}$B as a $^{17}$B-$n$-$n$ cluster, potential (1) was supplemented with a realistic $n$-$n$ interaction. The three-body problem was solved, using Faddeev equations in configuration space [1] and Gaussian Expansion method [22], and the energy of the $^{19}$B ground state provided by this three-body model was obtained.

The results are given in figure 4 (solid blue curve) as a function of the scattering length $a_S$. $^{19}$B appears to be bound in all the range of the experimentally allowed $a_S$ values, starting from $a_S \approx$ -30 fm. The binding energy increases with $\mid a_S \mid$ and saturates at the value $E_u = -0.081$ MeV in the limit $a_S \to -\infty$ which, by definition, is the unitary limit in the $n$-$^{17}$B channel (blue dotted points).

These results were obtained with an S-wave $n$-$n$ interaction adjusted to reproduce the experimental $a_{nn}$=-18.59 fm (see [20] for details). In order to study the full unitary limit of the model we have also set $a_{nn} \to -\infty$, by slightly modifying the attractive part of $V_{nn}$. The results of this limit correspond to the blue dashed horizontal line $E_{uu} = -0.160$ MeV.

We have also considered an alternative version of the three-body model by letting the $n$-$^{17}$B and the $n$-$n$ potentials to act in all partial waves. In this case we used the CI Bonn A model for the $n$-$n$ potential with $a_{nn} = -23.75$ fm. The results are indicated by black cuves on figure 4, including the unitary limit given now by $E_u = -0.18$ MeV. The differences come essentially from the differences in $a_{nn}$. P-wave contributions are small: for $a_S$=-150 fm they are ≈6 keV.

We would like to emphasize that, due to our ignorance of the $a_S$ value, we cannot predict a precise value for E($^{19}$B), but it is worth noticing that in all the domain of $a_S$ that we have considered, including the full unitary limit, the results provided by this simple model are compatible with the experimental value E=-0.14 ± 0.39 MeV.

In the right panel of figure 4 we have represented the modulus squared of the $^{19}$B ground state wave function as a function of the Jacobi coordinate $\mid \psi(r,R) \mid^2$ corresponding to $a_S$=-100 fm and $E=-0.130$ MeV. As a consequence of its weak binding energy, $^{19}$B is quite an extended dissymmetric object elongated in the $^{17}$B-$(nn)$ direction, as it corresponds to a two-neutron halo.

Apart from providing a good description of the $^{19}$B ground state, the model accommodates two broad 3-body resonances with total orbital angular momentum $L^{\pi}=1^-$ and $L^{\pi}=2^+$ respectively. Their parameters depend on $a_S$ and the model version. By fixing $a_S$=-150 fm, taking the interaction (1) in all partial waves and keeping S-wave interaction in $n$-$n$, their values are respectively $E_{L=1^-}$=+0.24(2)-0.31(4)i and $E_{L=2^+}$=+1.02(5)-1.22(6)i. Notice that the total angular momentum and parity $J^{\pi}$ of these resonant states results from the coupling between the quoted L and $J^{\pi}_{^{17}B}=3/2^-$, which are degenerated in our calculations. Several resonances in the continuum of $^{19}$B have been observed recently, although the determination of their energies and quantum number is still in progress [19]

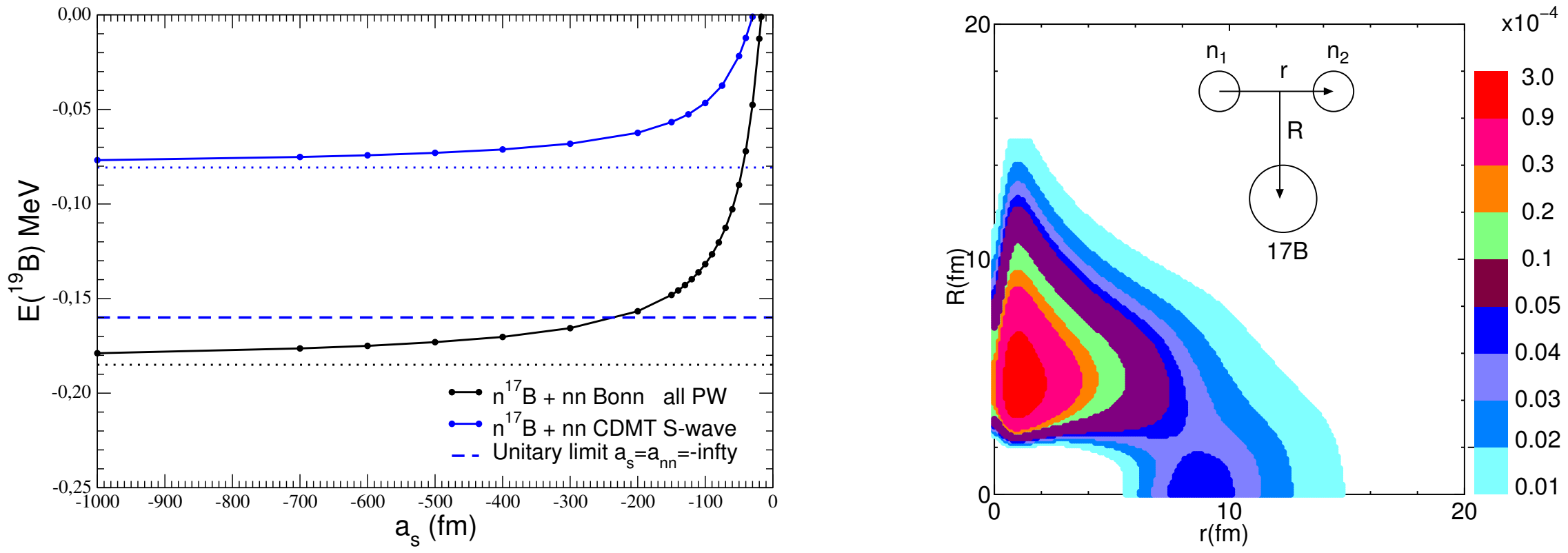

**Figure 4.** $^{19}$B ground-state energy with respect to the first particle threshold as a function of $a_S$, for $R=3$ fm (left panel) and (right panel) corresponding probability amplitude $|\Psi(r,R)|^2$ as a function of the Jacobi coordinates for $a_S=-100$ fm

As mentioned at the begining of this section, the results presented above neglect any spin-spin dependence in $V_{n^{17}\mathrm{B}}$. In particular they assume the equality of the $n$-$^{17}$B scattering length in the S=$1^-$ and S=$2^-$ channels: $a_{1^-}=a_{2^-}\equiv a_S$. In our previous work [20] we have introduced a spin-dependent interaction and checked the robustness of the $^{17}$B-$n$-$n$ model in what concerns the predictions of the $^{19}$B bound state: the $^{17}$B-$n$-$n$ ground state remains bound unless we introduce a spin-spin dependence at the level of $V_{n^{17}B}$ one order of magnitude greater than the one displayed in figure 1.

## 4. Conclusion

The strong repulsion observed in the low energy (S-wave) neutron-nucleus interaction for light nuclei A=2-6 starts becoming attractive in $^7$Li. When filling the P- and higher angular momentum neutron orbitals in the target, the Pauli repulsion among the incoming and the target neutrons weakens and the resulting balance with the attractive nN interaction (see figure 1) becomes attractive again.

In the case of $^{17}$B, this balance results into an extremely shallow virtual state which manifests by a huge scattering length $a_S$ <-50 fm, presumably in the total spin S=$2^-$ channel [18], experimentally observed but not yet precisely determined.

We have constructed a simple S-wave $n$-$^{17}$B interaction model and used it to describe the $^{19}$B isotope as a three-body $^{19}$B-$n$-$n$ cluster state. This model describes well the energy of the $^{19}$B ground state, in agreement with the measured binding energy, and accommodates two resonant state with total orbital angular momentum L=1 and L=2, also in agreement with experimental findings [19].

The success of this simple model lies on the double resonant character of the interaction both in the $n$-$^{17}$B and the $n$-$n$ channels which makes the $^{19}$B nucleus a nice illustration of a system described by the unitary limit of its interactions. Despite the large values of the scattering length in each channel, the system is still far to accommodate the first Efimov excited state, due to the existence of three different scattering length with only one being resonant and to the asymmetry among the constituent masses would require and $a_S$ value of few thousands fm.

The model that we have presented in this contribution can be naturally extended to describe the recently observed resonant states in $^{20}$B and $^{21}$B [23], by using the methods for solving the A=4 and A=5 systems that we developed in [24].

## Acknowledgments

We were granted access to the HPC resources of TGCC/IDRIS under the allocation 2018-A0030506006 made by GENCI (Grand Equipement National de Calcul Intensif). This work was supported by French IN2P3 for a theory project "Neutron-rich light unstable nuclei" and by the Japanese Grant-in-Aid for Scientific Research on Innovative Areas (No.18H05407).